\documentstyle[12pt]{article}
\newcommand{\ol}{\overline}
\newcommand{\ul}{\underline}
\newcommand{\bdm}{\begin{displaymath}}
\newcommand{\edm}{\end{displaymath}}
\newcommand{\nel}{\nu_{e}}
\newcommand{\nm}{\nu_{\mu}}
\newcommand{\nt}{\nu_{\tau}}
\newcommand{\na}{\nu_{1}}
\newcommand{\nb}{\nu_{2}}
\newcommand{\nc}{\nu_{3}}
\newcommand{\mf}{m_{\phi}}
\newcommand{\mg}{m_{G}}
\newcommand{\mr}{m_{R}}
\newcommand{\ml}{m_{L}}

\begin{document}
\begin{titlepage}
\setcounter{page}{0}
\rightline{Preprint YERPHI-1450(20)-94}

\vspace{2cm}
\begin{center}
{\Large THE NEUTRINO MASSES IN SUSY GUT}
\vspace{1cm}

{\large H. M. Asatrian, A. N. Ioannisian.} \\
\vspace{1cm}
{\em Yerevan Physics Institute, Alikhanyan Br. 2, Yerevan, Armenia}\\
{\em e-mail: "hrach@uniphi.yerphi.am"}\\
\end{center}

\vspace{5mm}
\centerline{{\bf{Abstract}}}
The neutrino mass problem in
$SU(4) \times SU(2)_{L} \times SU(2)_{R}$
SUSY GUT obtained from the compactification of $E_{8} \times E_{8}$
heterotic string is analyzed. The estimated values of the neutrino
masses and mixing angles
can explain the experimental data on solar neutrino flux.

\vfill
\centerline{\large Yerevan Physics Institute}
\centerline{\large Yerevan 1995}

\end{titlepage}
\newpage

     Recent  solar  neutrino  experiments  give  an  evidence  for
nonzero neutrino masses. The solar neutrino deficit can  be
explained in terms of the neutrino resonant oscillations if
the neutrino mass difference is of order:
$\Delta m^{2} \sim (0.3-1.2)\cdot10^{-5}eV^{2}$
or vacuum oscillations if
$\Delta m^{2} \sim (0.5-1.1)\cdot10^{-10}eV^{2}$
\cite{mikh,pont,blud}.

     It is well-known, that in $SO(10)$ GUT  small  neutrino  masses
can be obtained via seesaw  mechanism  \cite{gell}.  The  neutrino  mass
matrix for three  left
$\nu_{e}$, $\nu_{\mu}$,  $\nu_{\tau}$ and three right
 $\nu_{e}^{c}$,  $\nu_{\mu}^{c}$,  $\nu_{\tau}^{c}$
neutrinos has the following form:
\begin{equation}
\left( \begin{array}{cc}
0 & M_{D}\\                                
M_{D} & R
\end{array} \right)
\end{equation}

     In (1) M is a Dirac type $3 \times 3$  mass   matrix
(usually  it  is
equal to mass matrix of u, c, t, quarks), R- is the right neutrino
$3 \times 3$
mass matrix with entries much  greater  than  the  electroweak
symmetry breaking scale.  After  the  diagonalization  of  (1) one
obtains three  heavy  Majorana  states  (their  masses  practically
coincide with  the  eigenvalues  of  matrix  R)  and  three  light
Majorana states with the mass matrix:
\begin{equation}
M_{\nu}=\frac{M^{2}_{D}}{R}            
\end{equation}
     The scale  of matrix R entries can be of order of the  $SO(10)$
subgroup
$G = SU(4) \times SU(2)_{L} \times SU(2)_{R}$
breaking scale, if $G$   is broken by the  vacuum  expectation
value (v.e.v.) of Higgs field in the $\bf \ul{126}$  representation
of $SO(10)$.
For the superstring inspired models, however,  $G$                 can  be
broken only by the v.e.v. of Higgs field in $\bf \ul{16}$   representation
of $SO(10)$. In this case the masses of right neutrinos can arise only due
to the  radiative  corrections  \cite{witt80}.  For  the  SUSY  GUT,
however, this mechanism does not work \cite{ranpap}  and the right neutrinos
can obtain masses only due to the nonrenormalizable interactions  which
can arise in the superstring models \cite{papran}.

Our aim is consider the problem of the neutrino masses for the
supersymmetric model $G = SU(4) \times SU(2)_{L} \times SU(2)_{R}$,
which may (or may not) be considered as a subgroup of $SO(10)$.
The particle content of the model is the following \cite{antleo}:
the sixteen fermions for each generation (including the right neutrino)
belong to the representations {\bf F} and  $\bf \ol{F}$, where
\begin{eqnarray*}
F & = & (4,2,1)  = (u,d,\nu,e)\\
\ol{F} & = & (\ol{4}, 1, 2)
= (u^{c}, d^{c}, \nu^{c}, e^{c});
\end{eqnarray*}
and the Higgs fields- to the representations:
$\bf H=(4,1,2)$ and  $\bf \bar{H}=(\bar{4},1,2)$,
$\bf h=(1,2,2)$, $\bf D=(6,1,1)$.
The vacuum expectation values (v.e.v.) of the fields
H and $\ol{H}$ are connected with the breaking of the group $G$ and
the v.e.v. of the field  h    -with the breaking of the group
$G_{ew}=SU(2)_{L} \times U(1)_{Y}$ . In addition, there is a set of the
$G$  singlet scalar fields $ \phi_{m}$ (m=1,2...). For the
supersymmetric models, derived from the $E_{8} \times E_{8}$ heterotic
string compactification over Calabi-Yau manifolds with SU(3) holonomy,
the maximal gauge group in the four dimension is $E_{8} \times E_{6}$
(by embedding the spin connection of the manifold in the gauge
group, $E_{8} \times E_{8}$ can be broken to
$E_{8} \times E_{6}$ )
and chiral superfields belong to the
$\bf \ul{\ol{27}}, \ul{27}, \ul{1}$ representations of $E_{6}$
\cite{witt85}.In this case, the minimal set would consist of $n_{g}+1$
SO(10) (or G) singlets,
where $n_{g}$ is the number of fermion generations. One of the singlets
develops the v.e.v. at the electroweak scale and generates the masses of
the singlet fields \cite{antleo}.

Then most general superpotential for the supersymmetric
$G = SU(4) \times SU(2)_{L} \times SU(2)_{R}$ model
has the form  \cite{papran, antleo}  :
\begin{eqnarray}
\nonumber W & = & \lambda_{1}^{ij} F_{i} \bar{F_{j}} +
\lambda_{2}^{im} \bar{F_{i}} H \phi_{m} +                 
\lambda_{3} H H D + \\ & + & \lambda_{4} \bar{H} \bar{H} D +
\lambda_{5}^{m} \phi_{m} h h  +
\lambda_{6}^{mnl} \phi_{m} \phi_{n} \phi_{l} +\\
\nonumber & + & \lambda_{7}^{ij} F_{i}  F_{j} D  +
\lambda_{8}^{ij} \bar{F_{i}} \bar{F_{j}} +
\lambda_{9}^{m} D D \phi_{m}
\end{eqnarray}

The $9 \times 9$ mass matrix of the three left, three right neutrinos and
three singlets has the following form
\begin{equation}
\left( \begin{array}{ccc}
0 & M_{D} & 0 \\
M_{D} & R & M_{G} \\                        
0 & M_{G} & M_{\phi}
\end{array} \right),
\end{equation}
where $M_{G}$ is  $3 \times 3$  matrix of right neutrino-singlets mixing,
$M_{\phi}$ is $3 \times 3$  mass matrix of singlets. As a result one obtains
that  (in the case of absence of mixing) the light neutrino masses are
proportional to
$m_{q}^{2} m_{\phi}/m_{G}^{2}$, where $m_{q}^{2}$ for $q=u,c,t$
are the masses of u, c, t -quarks, $m_{\phi}$ is  the typical singlet
mass $\sim  M_{W}$ (electroweak breaking scale), $m_{G}$
is of order of G breaking scale $\sim  10^{16} GeV$  \cite{papran} .
This gives the ultralight neutrino masses: $m_{\nu_{1}} \sim 10^{-16}eV$,
$m_{\nu_{2}} \sim 10^{-12}eV$,
$m_{\nu_{3}} \sim 10^{-8}eV$. The similar problems exist in $SU(5)$
-flipped model. The various possibilities to obtain more
acceptable neutrino masses are considered in \cite{ranpap, papran, leonver}.
Here we want to consider the problem of the neutrino masses for the
superstring derived models proposed by Witten \cite{witt86}.
Witten \cite{witt86}
has shown that  it is possible to construct stable, irreducible
and holomorphic   $SU(4)$ or $SU(5)$ vector bundles over Calabi-Yau
manifolds.
This means that one can obtain an $SO(10)$ or $SU(5)$ supersymmetric gauge
theories  in four dimension by the embedding the structure group of the
bundle in $E_{8}$.  For the $SU(4)$
vector bundle, when the maximal gauge group in the four dimension
is SO(10), the content of the chiral superfields is
the following \cite{witt86,mur,asmur,asion}:
\begin{equation}
n_{g}\bf 16+\delta(\bf 16+\overline{\bf 16})+
\epsilon\bf 10+\eta \bf 1 ,                           
\end{equation}
where    $n_{g}$ is the number of  generations  ($n_{g}$=3), $\delta$,
$\epsilon$, $\eta$   are the
integer numbers $\delta,\epsilon, \eta \geq 1 $.
As for ordinary case of tangent bundle with the $E_{6}$ as a maximal
gauge group, in this case also it is possible to obtain models
with the gauge symmetries which are subgroups of SO(10) via
Hosotani mechanism
\cite{witt86,asmur,hos}.
Let us consider the neutrino mass problem in such a model with
gauge symmetry $G = SU(4) \times SU(2)_{L} \times SU(2)_{R}$.
For the simplest case  $\epsilon=\delta=1$ and $\eta=2$ one has only two
$SO(10)$  singlets. One of these singlets developed the  v.e.v. of
order of electroweak symmetry breaking scale. The second
singlet is mixed with the neutrinos. The   $7 \times 7$
mass matrix for  three left, three  right  neutrinos  and  this
singlet has the following form:
\begin{equation}
{\cal M}=\left( \begin{array}{rrr}
M_{L} & M_{D} & 0 \\
M_{D} & R & V \\                               
0 & V^{T} & m_{\phi}
\end{array} \right)
\end{equation}
where $M_{L}$, $M_{D}$ , R are $3 \times 3$ matrices, V-is a three
dimensional column. The elements of the left neutrino mass matrix $M_{L}$
can arise  due   to  the  nonrenormalizable  interactions
which are allowed in the string models \cite{akhm} :
\bdm
\lambda \frac{F F H H h h}{M^{3}},
\edm
where M is the typical scale connected with nonrenormalizable
interactions, it must be of order of Plank scale or string unification
scale, $\lambda $ -some constant.
Then the matrix elements of $M_{L}$  are  of order
 $\ml \sim  \frac{M_{W}^{2} \mg^{2}}{M^{3}}$.
     The 3x3 matrix R also arises  due  to  the  nonrenormalizable
interactions \cite{ranpap, papran}:
\bdm
\lambda '\frac{\bar{F} \bar{F} H H}{M},
\edm
The   matrix   elements
of R  are  of  order  $\mr \sim  \frac{\mg^{2}}{M}$.
The 3x3 neutrino Dirac mass matrix $M_{D}$
(which we assume  to be equal to the up quark  mass  matrix),
$V_{i}$  (i=1,2,3)- the mixing between  singlet   and
right neutrinos and the mass $\mf$ of the
singlet arise from the usual interaction terms (3).

To obtain the estimates for the neutrino masses one must determine
G breaking scale $\mg$. We will consider two cases: the
symmetry $G = SU(4) \times SU(2)_{L} \times SU(2)_{R}$ with and
without the left-right discrete symmetry.
The renormalization group
equations for one loop gives the following solutions for the coupling
constants $\alpha_{1}, \alpha_{2},\alpha_{3}$
\cite{asmur}:
\begin{equation}
\alpha_{i}^{-1}(M_{Z}) = \alpha_{X}^{-1} - \frac{b_{i}}{2\pi}
ln\frac{M_{S}}{M_{Z}} - \frac{b_{is}^{-}}{2\pi}
ln\frac{M_{R}}{M_{S}} - \frac{\hat{b}_{is}}{2\pi}                 
ln\frac{M_{X}}{M_{R}}
\end{equation}
where $M_{S}$ is the supersymmetry breaking scale (we assume that it
is between 100GeV and 10000GeV), the values of $b_{i}, b_{is}^{-},
\hat{b}_{is}$ (i=1,2,3) are given in \cite{asmur}. The more precise
results can be obtained in two loop approximation.
The renormalization group equations analysis in
two loop approximation gives the following results for $\mg$
\begin{equation}
\mg \sim (1.6\cdot10^{16} - 2.2\cdot10^{17})GeV       
\end{equation}
for the case of the presence of the discrete left-right symmetry and
\begin{equation}
\mg \sim (1.5\cdot10^{15} - 2\cdot10^{16})GeV         
\end{equation}
for the case of absence of such a symmetry. The results of (8) and (9)
are obtained for the initial values of the electroweak coupling constants
\begin{eqnarray}
\nonumber \alpha_{3}(M_{Z}) = 0.120 \pm 0.006 \\
sin^{2}\theta_{W} = 0.2328 \pm 0.0007 \\          
\nonumber \alpha^{-1}(M_{Z}) = 128.8 \pm 0.9
\end{eqnarray}

Let  as  consider  now  the  mass  matrix  (6).   After   the
diagonalization of the matrix (6) we obtain  7 Majorana  neutrinos
(three of which are light). One can estimate  the  light  neutrino
mass values without specifying the exact form of matrix  (6).  The
only assumption we made, is the following: all the matrix elements
of the matrix $M_{L}$  are of the same  order  of  magnitude
$\ml \sim  \frac{M_{W}^{2} \mg^{2}}{M^{3}}$.
The  same
statement  must  be  valid  for   the $3 \times 3$ matrix
  R   and  three dimensional column V separately: the elements of R
are of order $\mr \sim  \frac{\mg^{2}}{M}$ and the elements of V are
of order of $\mg$.
Then, it is  easy  to  estimate  the  value  of  the
determinant of the matrix (6) and the sums of its diagonal  minors
of 6x6, 5x5, 4x4 order:
\begin{eqnarray}
\nonumber det{\cal M} & \sim &
m_{c}^{2}m_{t}^{2}\frac{M_{W}^{2}\mg^{4}}{M^{3}}  \\
det{\cal M}_{6} & \sim & m_{c}^{2}m_{t}^{2}\mg^{2}  \\           
\nonumber det{\cal M}_{5} & \sim & m_{t}^{2}\frac{\mg^{4}}{M} \\
\nonumber det{\cal M}_{4} & \sim & \frac{\mg^{6}}{M^{2}}
\end{eqnarray}

Two eigenvalues of  the matrix
(6) are  of order $\mr$  and the other two- of order $\mg$.
One can obtain a simple formulae for the
masses of three light neutrinos for the case $m_{\nc} \gg
m_{\nb} \gg m_{\na}$ :
\begin{eqnarray}
\nonumber m_{\na} \sim \frac{det{\cal M}}{det{\cal M}_{6}}
\sim \frac{M_{W}^{2}\mg^{2}}{M^{3}}  \\
m_{\nb} \sim \frac{det{\cal M}_{6}}{det{\cal M}_{5}}
\sim r_{c}^{-1}\frac{m_{c}^{2}M}{\mg^{2}}  \\                          
\nonumber m_{\nc} \sim \frac{det{\cal M}_{5}}{det{\cal M}_{4}}
\sim r_{t}^{-1}\frac{m_{t}^{2}M}{\mg^{2}}
\end{eqnarray}
     In (11), (12) $r_{c}, r_{t}$ -are the factors connected  with  the
quark mass renormalization from the unification  scale to ordinary
energies. These factors depend on the ratio of  vacuum  expectation
values of scalar doublets connected with the electroweak  symmetry
breaking ($tg\theta$) and the t-quark mass. For the t-quark mass
$m_{t}=(175 \pm 15)$GeV and $0.1<tg\theta<10$ one obtains:
\begin{eqnarray}
\nonumber 20 > r_{c} > 3    \\
7 > r_{t} > 1
\end{eqnarray}                             
     In (13) the larger smaller of $r_{c}, r_{t}$  correspond
to  the larger values of $m_{t}$ .

     What about the mixing angles? To obtain the estimates for the
mixing  angles  between  the left $\nel, \nm, \nt$  neutrinos one has  to
specify the form of Dirac -type mass matrix $M_{D}$ .  Let  us  consider
the case when the mass  matrix  of  u,  c,  t-  quarks  (which  we
consider to be equal to $M_{D}$) has  a  form,  proposed  by  Fritzsch
\cite{fri}:
\begin{equation}
M_{D} = \left( \begin{array}{ccc}
0 & a & 0 \\
a & 0 & b \\                               
0 & b & c
\end{array} \right)
\end{equation}
where $a=\sqrt{m_{u}m_{c}}, b=\sqrt{m_{c}m_{t}}, c=m_{t}$.
We must made some assumptions for the matrix R and
$M_{L}$ also: let us consider the
case when they have a diagonal form:
\begin{equation}
R= \left( \begin{array}{ccc}
R_{1} & 0 & 0 \\
0 & R_{2} & 0 \\                               
0 & 0 & R_{3}
\end{array} \right)
\hspace{0.5cm}
M_{L} = \left( \begin{array}{ccc}
m_{1} & 0 & 0 \\
0 & m_{2} & 0 \\
0 & 0 & m_{3}
\end{array} \right)
\end{equation}
     It is possible  to estimate the  mixing  angles  for  three
light  neutrino  states  $\na, \nb, \nc$   by  solving  the  equation  for
eigenvectors and eigenvalues of the matrix ${\cal M}$
for the case $m_{\nc} \gg m_{\na}, m_{\nb}$. Using
the following formulae for the light neutrino masses:
\begin{eqnarray}
\nonumber
m_{\nc}  \simeq \frac{det{\cal M}_{5}}{det{\cal M}_{4}}  \\
m_{\nb}m_{\na} \simeq \frac{det{\cal M}}{det{\cal M}_{5}}  \\  
\nonumber
m_{\nb} + m_{\na} \simeq \frac{det{\cal M}_{6}}{det{\cal M}_{5}}
\end{eqnarray}
and calculating the determinant of matrix ${\cal M}$ and sums of its
main minors of 6, 5, and 4 order
\begin{eqnarray}
\nonumber
det{\cal M} & \simeq & -m_{1}m_{c}^{2}m_{t}^{2}M_{1}^{2} +
m_{1}m_{2}m_{t}^{2}
(R_{1}M_{2}^{2}+R_{2}M_{1}^{2}) \\
\nonumber
& & \mbox{} -2m_{1}\sqrt{m_{u}m_{c}}m_{c}m_{t}^{2}M_{1}M_{2}
-2m_{1}m_{2}\sqrt{m_{t}m_{c}}m_{t}R_{1}M_{2}M_{3}\\
\nonumber
& & \mbox{} -m_{1}m_{u}m_{c}m_{t}^{2}M_{2}^{2}
-m_{2}m_{u}m_{c}m_{t}^{2}M_{2}^{2} \\
det{\cal M}_{6} & \simeq & -m_{c}^{2}m_{t}^{2}M_{1}^{2} +
(m_{1}+m_{2})m_{t}^{2}
(R_{1}M_{2}^{2}+R_{2}M_{1}^{2}) \\
\nonumber
& & \mbox{} -2\sqrt{m_{u}m_{c}}m_{c}m_{t}^{2}M_{1}M_{2}
-2(m_{1}+m_{2})\sqrt{m_{t}m_{c}}m_{t}R_{1}M_{2}M_{3}\\
\nonumber
det{\cal M}_{5} & \simeq & m_{t}^{2}
(R_{1}M_{2}^{2}+R_{2}M_{1}^{2})
-2\sqrt{m_{c}m_{t}}m_{t}R_{1}M_{2}M_{3} \\                        
\nonumber
det{\cal M}_{4} & \simeq & M_{1}^{2}R_{2}R_{3}+M_{2}^{2}R_{1}R_{3}+
M_{3}^{2}R_{1}R_{2}
\end{eqnarray}
(where  $M_{1}, M_{2}, M_{3}$ are the elements in $\cal M$ connected with the
mixing of right neutrinos and singlet: $V^{T} = (M_{1}, M_{2}, M_{3})$)
one can obtain the estimates for the light neutrino masses. In (17) we
omit the terms not relevant for our consideration.
We will consider two alternatives for the $\na, \nb$:
\begin{eqnarray}
m_{\na} \ll m_{\nb}   \\                                             
m_{\na} \sim m_{\nb}                                                 
\end{eqnarray}
The conditions (18), (19) are equivalent to the conditions
\begin{eqnarray}
m_{G} \ll \sqrt{\frac{m_{c}}{M_{W}}}M   \\       
m_{G} \sim \sqrt{\frac{m_{c}}{M_{W}}}M                   
\end{eqnarray}
For the first case (18), (20) one obtains the same formulae for
neutrino masses as previously
\begin{equation}
m_{\na} \sim m_{L} \hspace{0.5cm} m_{\nb} \sim \frac{m_{c}^{2}}{r_{c}R}
\hspace{0.5cm} m_{\nc} \sim \frac{m_{t}^{2}}{r_{t}R}   
\end{equation}
where we assume that $R \sim R_{1} \sim R_{2} \sim R_{3}$
and $m_{1} \sim m_{2} \sim m_{3}$.
The three eigenstates $\nu_{i}$, i=1,2,3
of matrix (6) with three lightest masses
one can express via weak eigenstates $\nu_{\alpha}$, $\alpha=e,\mu,\tau$
by means of unitary transformation:
$\nu_{i} = a_{i\alpha}\nu_{\alpha}$,
where $a_{i\alpha}$ is the unitary matrix.
For the mixing angles between $\nel$ and $\nm$ ($\theta$)
and $\nel$ and $\nt$ ($\theta'$)
one obtains:
\begin{eqnarray}
\nonumber
tan\theta \sim \frac{a_{1\mu}}{a_{1e}} \sim \sqrt{\frac{m_{u}}{m_{c}}}\\
tan\theta' \sim \frac{a_{1\tau}}{a_{1e}} \sim \sqrt{\frac{m_{u}}{m_{t}}} 
\end{eqnarray}
Taking into account the results (8), (9) for the G symmetry breaking
scale one obtains the following estimates for the three light neutrino masses:
\begin{eqnarray}
\nonumber m_{\na} & \sim & (10^{-12} - 2\cdot10^{-10})
\left(\frac{M_{Pl}}{M}\right)^{3}eV \\
m_{\nb} & \sim & (4\cdot10^{-8} - 10^{-5})\frac{M}{M_{Pl}}eV \\         
\nonumber m_{\nc} & \sim & (8\cdot10^{-3} - 0.6)\frac{M}{M_{Pl}}eV
\end{eqnarray}
for case of the presence of discrete left-right (case (a)) symmetry and
\begin{eqnarray}
\nonumber m_{\na} & \sim & (10^{-14} - 10^{-12})
\left(\frac{M_{Pl}}{M}\right)^{3}eV \\
m_{\nb} & \sim & (6\cdot10^{-6} -2.5\cdot10^{-3})\frac{M}{M_{Pl}}eV \\  
\nonumber m_{\nc} & \sim & (0.3 - 150)\frac{M}{M_{Pl}}eV
\end{eqnarray}
for the case of the absence of left-right discrete symmetry (case (b)).
In (24), (25) $M_{Pl} = 1.2\cdot10^{19}GeV$ is the Plank mass.
Of course, all  the  estimates (24), (25)  are correct  with
accuracy  of order of  magnitude.

To explain the solar neutrino deficit via resonant oscillations
the neutrino mass difference and mixing angles must be of order
\begin{eqnarray}
\nonumber
\sqrt{\Delta m^{2}} \sim(1.7 - 3.5)10^{-3}eV \\
sin^{2}2\theta \sim (0.6-1.4)10^{-2} \hspace{0.5cm} or 
\hspace{0.5cm} (0.65-0.85)
\end{eqnarray}
Then for the case (a) it is possible to obtain
such a mass difference between third and second and first neutrinos if
the fraction $\frac{M}{M_{Pl}}$ is of order $\sim (1 - 0.01)$ which is
reasonable value \cite{asmur}.
The problem arises with mixing angle:
the formula (23) gives small value
for $sin^{2}2\theta \sim 10^{-4}$.

For the case (b) such a mass difference it is
possible to obtain between the second and the first neutrinos if the
fraction  $\frac{M}{M_{Pl}}$ is order of unity. In this case the condition
(20) is valid. The mixing angle is of order
$sin^{2}\theta \sim 1.3\cdot 10^{-2}$ which is a reasonable value.
Thus, in this case obtained values for the mixing angles  and neutrino
mass difference allow one to explain the solar neutrino deficit
via resonant oscillations in the sun.

To explain the solar neutrino deficit via longwave vacuum oscillations
the neutrino mass difference and the mixing angles must be of order
\begin{eqnarray}
\nonumber
\sqrt{\Delta m^{2}} \sim (0.5-1.1)10^{-5}eV \\
sin^{2}2\theta >0.75                                      
\end{eqnarray}
It is clear  that it is impossible for the considering case  as our mixing
angles are too small.

Let as proceed to the case (19), (21). For the case (a) the condition
(21) gives for M:
\begin{equation}
\frac{M}{M_{Pl}}=(0.018-0.22)               
\end{equation}

For such a values of M the light neutrino masses are of order
\begin{eqnarray}
\nonumber
m_{\na} \sim m_{\nb} & \sim & 4\cdot (10^{-8}-10^{-5})\frac{M}{M_{Pl}}eV
\sim (7.1\cdot 10^{-10}-2.2\cdot10^{-6})eV \\
m_{\nc} & \sim & (8\cdot10^{-3} - 0.6)\frac{M}{M_{Pl}}eV   
\sim (1.4\cdot 10^{-4}-0.13)eV
\end{eqnarray}
It is clear that the mass difference of second and first family is
too small. The mass difference between third and first family can
achieve acceptable value
but the mixing angle as in previous case
will be small $sin^{2}2\theta \sim \cdot 10^{-4}$.
Let us consider the case (b). From (21) one obtains for M
\begin{equation}
\frac{M}{M_{Pl}}=(0.0014-0.02)               
\end{equation}
For such a values of M the light neutrino masses are of order
\begin{eqnarray}
\nonumber
m_{\na} \sim m_{\nb} & \sim & (6\cdot 10^{-6}-2.5\cdot 10^{-3})
\frac{M}{M_{Pl}}eV
\sim (8.4\cdot 10^{-9}-5\cdot10^{-5})eV \\
m_{\nc} & \sim & (0.3 - 150)\frac{M}{M_{Pl}}eV     
\sim (4.2\cdot 10^{-3}-3.0)eV
\end{eqnarray}
The value $(0.5-1.1)10^{-5}$eV which is necessary to explain
the solar neutrino deficit via longwave vacuum oscillations
is achieved for the mass difference of first two neutrinos
for the following values of M:
\begin{equation}
\frac{M}{M_{Pl}}=(0.02-0.002)              
\end{equation}
What about the mixing angles for  the considering case?
As the masses of two lightest neutrinos
are close to each other, the mixing angle between them can be relatively
large.

The angle relevant for the neutrino
oscillations is given by (if we neglect the mixing with third light
neutrino):
\begin{equation}
\tan \theta = \frac{a_{1\mu}}{a_{1e}} \simeq
\sqrt{\frac{m_{c}}{m_{u}}}\frac{m_{\na}-m_{1}}{m_{\na}-m_{2}}    
\end{equation}
where
\begin{eqnarray}
\nonumber
m_{\na} & = & m_{1} + \frac{1}{2}m
\left (1+\frac{m_{1}-m_{2}+m'}{m_{1}-m_{2}-m'}\right )  \\
\nonumber
m_{\nb} & = & m_{2} + m'+ \frac{1}{2}m
\left (1-\frac{m_{1}-m_{2}+m'}{m_{1}-m_{2}-m'}\right )  \\         
m & = & -\frac{m_{u}m_{c}m_{t}^{2}M_{1}M_{2}}{det{\cal M}_{5}}\\
\nonumber
m' & = & -\frac{m_{c}^{2}m_{t}^{2}M_{1}^{2}+m_{u}m_{c}m_{t}^{2}M_{2}^{2}+
2\sqrt{m_{u}m_{c}}m_{c}m_{t}^{2}M_{1}M_{2}}{det{\cal M}_{5}}
\end{eqnarray}
To explain the solar neutrino deficit via vacuum longwave oscillations
$tan\theta$ must satisfy the condition
\begin{equation}
0.58<tan\theta<1.75                   
\end{equation}
(as it follows from (27)).
This gives :
\begin{equation}
(m_{1}-m_{2}-m')^{2} \sim mm'        
\end{equation}
or, with the same accuracy:
\begin{equation}
(m_{\na}-m_{\nb})^{2} \sim mm'        
\end{equation}

Thus, it is possible to obtain large (35) mixing angles in this
case if the condition (36) takes place. This means that the
masses of first two light neutrinos must be very close to each other: the
difference of masses must be $\sim$ 100 times  smaller than the masses.

Thus, our analyze shows that it is possible to explain the solar
neutrino deficit for the considering theory for the case of a
unification group $G = SU(4) \times SU(2)_{L} \times SU(2)_{R}$ without
the left-right discrete symmetry can be explained in two ways.
1)Via resonant
oscillations in the sun if the value of which is characteristic
scale of nonrenormalizable interactions is of order of Plank mass
$M_{Pl}=1.2\cdot 10^{19}GeV$.
Generally speaking, it is more natural, that such a interactions
arises in string scale
$M_{s}=\frac{g_{string}M_{Pl}}{\sqrt{8\pi}}=1.7\cdot 10^{18}GeV$
\cite{wain,kapl} however
the difference of $\sim 7$ can be explained by means of coupling constants
$\lambda$, $\lambda '$ in nonrenormalizable terms:
\bdm
\lambda \frac{F F H H h h}{M^{3}},
\hspace{0.5cm}
\lambda '\frac{\bar{F} \bar{F} H H}{M}
\edm
The mixing angle in this case $\sim 0.06$ which is the value we need.
2)Via longwave vacuum oscillations: in this case the masses of two
light neutrinos are very close to each other and the mixing angle is
large as needed.

Authors thank H. H. Asatryan for help in most of calculations.

    The research described in this publication was made possible in part
by Grant N MVU000 from the International Science Foundation.

\end{document}